\documentclass[a4paper]{article}
\usepackage{graphicx}
\usepackage{mathptmx}
\usepackage{amsmath}
\usepackage{amsfonts}
\usepackage{amssymb}
\usepackage{multirow}
\usepackage[]{units}
\usepackage[]{nicefrac}
\usepackage{microtype}

\usepackage{natbib}

\usepackage{authblk}

\bibpunct{(}{)}{;}{a}{,}{,}

\title{The Green's function formalism as a bridge between single and multi-compartmental modeling}

\author[1]{Willem A.M. Wybo\thanks{willem.wybo@epfl.ch}}
\author[2]{Klaus M. Stiefel\thanks{k.stiefel@uws.edu.au}}
\author[1]{Benjamin Torben-Nielsen\thanks{btorbennielsen@gmail.com}}
\affil[1]{Blue Brain Project, Swiss Federal Institute of Technology in Lausanne (EPFL), Switzerland}
\affil[2]{\textsc{marcs} Institute, University of Western Sydney \\ 						Sydney, Australia}

\date{}

\begin{document}

\maketitle

\begin{abstract}
Neurons are spatially extended structures that receive and process inputs on their dendrites. It is generally accepted that neuronal computations arise from the active integration of synaptic inputs along a dendrite between the input location and the location of spike generation in the axon initial segment. However, many application such as simulations of brain networks, use point-neurons --neurons without a morphological component-- as computational units to keep the conceptual complexity and computational costs low. Inevitably, these applications thus omit a fundamental property of neuronal computation. In this work, we present an approach to model an artificial synapse that mimics dendritic processing without the need to explicitly simulate dendritic dynamics. The model synapse employs an analytic solution for the cable equation to compute the neuron's membrane potential following dendritic inputs. Green's function formalism is used to derive the closed version of the cable equation. We show that by using this synapse model, point-neurons can achieve results that were previously limited to the realms of multi-compartmental models. Moreover, a computational advantage is achieved when only a small number of  simulated synapses impinge on a morphologically elaborate neuron. Opportunities and limitations are discussed. 
\end{abstract}

\section{Introduction}
\label{sec:introduction}

Neurons are morphological structures: they have dendritic branches on which most inputs are received and an axonal tree through which the output signal is communicated with other neurons. In this light, neuronal computations can be seen as the integration of synaptic inputs along the dendrites up to the axon initial segment where an output signal is generated. Hence a key role in neuronal computation is taken by the exact shape and composition of dendrites. Indeed, it is known that the neuronal response is shaped by the precise location and activation pattern of synapses \citep{Branco2010, Torben-Nielsen2010, Gidon2012} and by the expression and distribution of (voltage-gated) ion-channels \citep{Migliore2002, Magee1999, Torben-Nielsen2010,Spruston2008}. 

Despite this proven importance, dendritic processing is usually ignored in network simulation \citep{Gewaltig2007,Brette2007,Richert2011}, but see \citep{Markram2006} for an exception. One reason is the computational cost associated with multi-compartmental simulations: a costs that, at the level of the model neuron, scales with the morphological complexity of the dendritic arborization. Related is the conceptual cost associated with building detailed single-neuron models \citep{Hay2013} with the spatial distribution of conductances across the membrane and localized non-linearities. The key is to capture the somatic voltage in response to synaptic inputs on the dendrites. Is there an alternative to multi-compartmental models to simulate the effects of dendrites on synaptic potentials, without large computational overhead?

To this end, two strategies are commonly adopted in the literature. The first consist of performing a morphological reduction by reducing the number of dendritic segments while attempting to capture crucial characteristics of dendritic processing \citep{Traub2005,Kellems2010}. A second strategy is to by-pass multiple (dendritic) compartments altogether by using point-neurons and fit voltage-kernels that matches the dendritic signal transformation shaping the voltage waveform caused by a synaptic input at the soma \citep{Jolivet2004, Gutig2006}. The fitted \citep{Jolivet2004} or learned \citep{Gutig2006} kernel is then simply added to the somatic membrane potential. While this strategy is computationally efficient and some temporal effects of dendritic processing can be captured, it is a rather crude approximation of what dendritic integration stands for and elementary features of dendritic processing, such as local interaction between inputs, are impossible to achieve. 

In this work we present a true alternative based on applying the Green's function formalism to cable theory. This way we can exactly compute the effect of synaptic inputs located in the dendrites on the somatic membrane potential \citep{Koch1985}. By design we thus compute the linear transfer function between the site of the synaptic inputs and the soma. The main advantage of this approach is that the effect of synaptic inputs along a dendrite on the somatic membrane potential can be calculated analytically. Consequently, simulations in our model are independent of the morphological complexity and a full reduction to a point-neuron can be used, as the entire effect of the morphology is captured in a transfer function. This property sets our approach apart from existing methods to model dendrites implicitly: the approach based on the equivalent cable works only with geometrically tightly constrained morphologies \citep{Ohme1998}, while, as in \citep{VanPelt1992} all branch points of a dendritic tree have to be modeled explicitly. Because we capture arbitrary dendritic morphologies by means of transfer functions, our synapse model is able to use dendrite-specific mechanism of computation, such as delay lines (as \citep{Gutig2006}) but also local non-linearities due to membrane saturation. Hence, we can capture fundamental features of dendritic integration by directly deriving the Green's function from dendritic cable theory. 

We implemented our synapse model in the Python programming language as a proof of principle, and validated it by evaluating its correctness and execution times on two tasks. First, we show that a morphology-less point-neuron equipped with the proposed synapse model can exploit differential dendritic processing to perform an input-order detection task \citep{Agmon-Snir1998}. We show that both for passive models and models with active currents in the soma, the agreement with a reference \textsc{neuron} simulation \citep{Carnevale2006} is seamless. Second, we show that the proposed neuron model is capable of accurate temporal integration of multiple synaptic inputs, a result for which knowledge of the precise neuronal morphology in relation to the synaptic locations is imperative. To this end, we construct a point-neuron model mimicking the dendritic processing in the dendrites of a Layer 5 pyramidal cell. Again, we demonstrate that the agreement with a reference \textsc{neuron} simulation is seamless. By providing this example, we demonstrate that our proposed approach is highly suitable for the common scenarios to investigate dendritic processing. In such scenarios, the somatic response to a limited number of synapses located in the dendrites is measured while changing the dendritic properties.

\section{Synapse model based on the Green's function formalism}\label{sec:methods}

The core rationale of this work is the simplification of a passive neuron model by analytically computing the transfer function between synapses and the soma. Solving the cable equation for dendrites is not new, and several ways are documented \citep{Koch1985, Butz1974, Norman1972}. The application of the cable equation to simplify arbitrarily morphologically extended multi-compartmental models to a point-neuron is, however, new.

By solving the cable equation, we thus substitute the effects of an electrical waveform traveling down a dendrite by a so-called pulse-response kernel. Conceptually, we think of the neural response to a spike input as being characterized by three functions: the conductance profile of the synapse, the pulse-response kernel at the synapse and the pulse-response transfer kernel between the input location and the soma to mimic the actual dendritic propagation. The first function is chosen by the modeller: common examples are the alpha function, the double exponential or the single decaying exponential \citep{Rotter1999,Giugliano2000,Carnevale2006}. The second function captures the decay of the voltage at the synapse given a pulse input, and thus allows for a computation of the synaptic driving force, whereas the third function allows for the computation of the response at the soma, given the synaptic profile, driving force, and dendritic profile.

More formally, we write $g(t)$ for the synaptic conductance profile, $G_{\text{syn}}(t)$ for the pulse response kernel at the synapse and $G_{\text{som}}(t)$ for the pulse response kernel between synapse and soma. Then, given a presynaptic spiketrain $\{ t_s \}$ and a synaptic reversal potential $E_r$, the somatic response of the neuron is characterized by:
\begin{eqnarray}\label{eq:intro}
\begin{aligned}
g(t) & = F(\mathbf{a}(t)), \hspace{4mm} \frac{\mathrm{d}\mathbf{a}}{\mathrm{d}t}(t) = H(\mathbf{a}(t),\{t_s\})\\
V_\text{syn}(t) & = \int_{-\infty}^{t} \mathrm{d}k \ G_{\text{syn}}(t-k) \ g(k) \ (V_\text{syn}(k)-E_r) \\
V_\text{som}(t) & = \int_{-\infty}^{t} \mathrm{d}k \ G_{\text{som}}(t-k) \ g(k) \ (V_\text{syn}(k)-E_r),
\end{aligned}
\end{eqnarray}
where $E_r$ is the synaptic reversal potential, $F(.)$ and $H(.)$ depend on the type of synapse chosen and $\mathbf{a}$ denotes the set of synaptic parameters required to generate the conductance profile $g(t)$. Our task is to compute  $G_{\text{syn}}(t)$ and $G_{\text{som}}(t)$. We will show that these functions follow from the Green's function formalism. 

\subsection{The neuron model in time and frequency domains}

\subsubsection{Time domain}

Here, we assume a morphological neuron models with passive dendritic segments. Each segment, labeled $d = 1,\hdots,N$, is modeled as a passive cylinder of constant radius $a_d$ and length $L_d$. It is assumed that all segments have an equal membrane conductance $g_m$, reversal potential $E$, intracellular axial resistance $r_a$ and membrane capacitance $c_m$. By convention we label the locations along a dendrite by $x$, with $x=0$ and $x=L_d$ denoting the proximal and distal end of the dendrite, respectively. Then, in accordance with cable theory, the voltage in a segment $d$ follows from solving the partial differential equation \citep{Tuckwell1988Introduction}:
\begin{eqnarray}\label{eq:cable}
\begin{aligned}
\frac{\pi a_d^2}{r_a}\frac{\mathrm{\partial}^2 V_d}{\mathrm{\partial}x^2}(x,t) \ - \ 2\pi a_d g_m V_d(x,t) \ - 2\pi a_d c_m \frac{\mathrm{\partial} V_d}{\mathrm{\partial}t}(x,t) \ = \ I_d(x,t),
\end{aligned}
\end{eqnarray}
where $I_d(x,t)$ represents the input current in branch $d$, at time $t$ and at location $x$. We assume that the dendritic segments are linked together by boundary conditions that follow from the requirement that the membrane potential is continuous and the longitudinal currents (denoted by $I_{ld}$) conserved:
\begin{eqnarray}
\begin{aligned}
V_{d}(L_{d}, t) & = V_{i}(0,t), \hspace{4mm} i \in \mathcal{C}(d) \\
I_{ld}(L_{d}, t) & = \sum_{i \in \mathcal{C}(d)} I_{li}(0,t)
\end{aligned}
\end{eqnarray}
where $\mathcal{C}(d)$ denotes the set of all child segments of segment $d$. The longitudonal currents are given by:
\begin{equation}
I_{ld}(x,t) = \frac{\pi a_d^2}{r_a}\frac{\mathrm{\partial} V_d}{\mathrm{\partial}x}(x,t). 
\end{equation}
Different dendritic branches originating at the soma are joined together by the lumped-soma boundary condition, which implies for the somatic voltage $V_{\text{som}}(t)$:
\begin{equation} \label{eq:lsb1}
V_{\text{som}}(t) = V_d(0,t) \hspace{3mm} \forall d \in \mathcal{C}(\text{soma})
\end{equation}
and
\begin{equation}\label{eq:lsb2}
\sum_{d=1}^{\mathcal{C}(\text{soma})} I_{ld}(0,t) = I_{\text{som}}(V_{\text{som}}(t)) + C_{\text{som}} \frac{\mathrm{\partial} V_{\text{som}}}{\mathrm{\partial}t}(t),
\end{equation}
with $I_{\text{som}}$ denoting the transmembrane currents in the soma, that can be either passive or active. Note that, for all further calculations, we will treat $I_{\text{som}}(V_{\text{som}}(t))$ as an external input current, and apply the Green's function formalism only on a soma with a capacitive current.
For segments that have no children (i.e., the leafs of the tree structure), the sealed end boundary condition is used at the distal end:
\begin{equation}
I_{ld}(L_d,t) = 0 \hspace{3mm} \forall d.
\end{equation}

\subsubsection{Frequency domain}

Fourrier-transforming this system of equations allows for the time-derivatives to be written as complex multiplications, for which analytic \citep{Butz1974} or semi-analytic \citep{Koch1985} solutions can be computed. Doing so transforms equation \eqref{eq:cable} into:
\begin{equation}\label{eq:freqcable}
\frac{\mathrm{\partial}^2 V_d}{\mathrm{\partial}x^2}(0,\omega) - \gamma _d(\omega)^2 V_d(x,\omega) = I_d(x,\omega)
\end{equation}
where $\omega$ is now a complex number and $\gamma_d(\omega)$ is the frequency-dependent space constant, given by
\begin{equation}
\gamma_d(\omega) = \sqrt{\frac{z_{ad}}{z_{md}(\omega)}}
\end{equation}
with $z_{ad} = \frac{r_a}{\pi a_d^2}$ the dendritic axial impedance and $z_{md} = \frac{1}{2 \pi a_d (i c_m \omega + g_m)}$ the membrane impedance in branch $d$. The lumped soma boundary conditions \eqref{eq:lsb1} and \eqref{eq:lsb2} become
\begin{equation} 
V_{\text{som}}(\omega) = V_d(0,\omega) \hspace{3mm} \forall d
\end{equation}
and
\begin{equation}
\sum_{d=1}^N I_{ld}(0,\omega) = \sum_{d=1}^N \frac{1}{z_{ad}}\frac{\mathrm{\partial} V_d}{\mathrm{\partial}x}(0,\omega) = \frac{1}{Z_{\text{som}}(\omega)}V_{\text{som}}(\omega),
\end{equation}
where
\begin{equation}
Z_{\text{som}}(\omega) = \frac{1}{i C_{\text{som}} \omega}
\end{equation}
is the somatic impedance. The sealed-end boundary conditions are:
\begin{equation}\label{eq:bcfreq}
I_{ld}(L_d,\omega) = \frac{1}{Z_L} V_d(L_d,\omega) = 0
\end{equation} 
with sealed-end impedance $Z_L = \infty$.

\subsection{Morphological simplification by applying Green's function}
Here we will describe the Green's function formalism formally in the time domain to explain the main principles. In the next paragraph we will then turn back to the frequency-domain to compute the actual solution. For the argument we consider a general current input $I_d(x,t)$. In the case of dynamic synapses, such a current input is obtained from the synaptic conductances by the Ohmic relation:
\begin{equation}\label{eq:current}
I_{d}(x,t) = g(t)(E_r-V_d(x,t))
\end{equation}
or, in the case of active channels, from the ion channel dynamics
The cable equation \eqref{eq:cable} can be written formally as: 
\begin{equation}\label{eq:operator}
\hat{L}_d V_d(x,t) = I_d(x,t)
\end{equation}
where $\hat{L}_d = \frac{\pi a_d^2}{r_a}\frac{\mathrm{\partial}^2 }{\mathrm{\partial}x^2} - 2\pi a_d g_m - 2\pi a_d c_m \frac{\mathrm{\partial}}{\mathrm{\partial}t}$ is a linear operator\footnote{Note that formally, the operator $\hat{L}_d$ depends on $x$ explicitly in a discontinuous way: for  for $0<x<L_d$: $\hat{L}_d(x) = \frac{\pi a_d^2}{r_a}\frac{\mathrm{\partial}^2 }{\mathrm{\partial}x^2} - 2\pi a_d g_m - 2\pi a_d c_m \frac{\mathrm{\partial}}{\mathrm{\partial}t}$, and for $x=0$: $\hat{L}_d(x=0) =  \sum_{d=1}^N \frac{\pi a_d^2}{r_a}\frac{\mathrm{\partial}}{\mathrm{\partial}x} - G_{\text{som}} - C_{\text{som}} \frac{\mathrm{\partial}}{\mathrm{\partial}t}$.}, which means that for two arbitrary functions $V_1(x,t)$ and $V_2(x,t)$ the following identity holds:
\begin{equation} \label{eq:linearity}
\hat{L}_d (aV_1(x,t) + bV_2(x,t)) = a \hat{L}_d V_1(x,t) + b \hat{L}_d V_2(x,t)
\end{equation}
The Green's function of the system is then defined as the solution of the following differential equation:
\begin{equation}
\hat{L}_d G_{dd'}(x,x',t,t') = \delta(x-x')\delta(t-t')\delta_{dd'}.
\end{equation}
which also justifies its name as ``pulse-response kernel''. The solution to the general input current $I_d(x,t)$ is then written as 
\begin{equation}\label{eq:greengeneral}
V_d(x,t) = \sum_{d'} \int_0^{L_d} \mathrm{d}x' \int_{-\infty}^t \mathrm{d}t' G_{dd'}(x,x',t,t') I_{d'}(x',t'),
\end{equation}
which can be verified by substituting this equation in \eqref{eq:operator} and using the assumption of linearity \eqref{eq:linearity}.

Two considerations allow us to simplify this system: first, as a consequence of the fact that the operator $\hat{L}_d$ is translation invariant in the time domain, the Green's function only depends on temporal differences:
\begin{equation}
G_{dd'}(x,x',t,t') = G_{dd'}(x,x',t-t'),
\end{equation}
second, in the case of neuronal dynamics, it often suffices to consider inputs at a discrete number of locations, labeled $x_i^{d'}$, with $d'$ denoting the segment of the input location:
\begin{equation}
I_{d}(x,t) = \sum_{x_i^{d'}} I_{d}(x,t)\delta(x-x_i^{d'})\delta_{dd'},
\end{equation}
where $I_{d}$ can either denote a synaptic input current or an active membrane current at a point-like location (here we only consider active currents at the soma). Given these considerations, equation \eqref{eq:greengeneral} reduces to 
\begin{equation}\label{eq:greensimple}
\begin{aligned}
V_d(x,t) =\sum_{x_i^{d'}} \int \mathrm{d}t' G_{dd'}(x,x_i^{d'},t-t') I_{d'}(x_i^{d'},t').
\end{aligned}
\end{equation}

Then it follows from equations \eqref{eq:current} and \eqref{eq:greensimple} the membrane potentials at the $N$ synapses (labeled $i$) distributed on dendritic branches $d_i$ at locations $x_1^{d_1},\hdots,x_n^{d_n}$ are
\begin{equation}\label{eq:greenssynapse}
\begin{aligned}
& V_d(x_j^{d_j},t) = \\
& \hspace{4mm} \sum_{i} \int_{-\infty}^t \mathrm{d}t' \ G_{d_j d_i}(x_j^{d_j},x_i^{d_i},t-t') \ g_i(t') \ (E_r-V_{d_i}(x_i^{d_i},t')) + \\
 & \hspace{4mm} \int_{-\infty}^t \mathrm{d}t' G_{d_i d_i}(x_j^{d_i},\text{soma},t-t') I_{\text{som}}(V_{\text{som}}(t')),
\end{aligned}
\end{equation}
whereas the potential at the soma is given by:
\begin{equation}\label{eq:greenssoma}
\begin{aligned}
& V_{\text{som}}(t) = \\
& \hspace{4mm} \sum_{i} \int_{-\infty}^t  \mathrm{d}t' \ G_{d_i d_i}(\text{soma},x_i^{d_i},t-t') \ g_i(t') \ (E_r-V_d(x_i^{d_i},t')) + \\
& \hspace{4mm} \int_{-\infty}^t \mathrm{d}t' G(\text{soma},\text{soma},t-t') I_{\text{som}}(V_{\text{som}}(t')).
\end{aligned}
\end{equation}

\subsubsection{Frequency domain solution for the Green's function}

Let us now turn to the calculation of the Green's function for a pulse input at time $t=0$ and at a location $x_i$ in dendrite $d$. Here, we perform this calculation in the frequency domain, whereas in the next paragraph we will show how the inverse transform, back to the time domain, can be evaluated. To that end we use the algorithm described in \citep{Koch1985}. As an example, we describe this procedure for a simplified morphology, where each dendritic branch arriving at the soma is modeled as a single cylinder (we use this morphology in section \ref{sec:iodetect} on input-order detection). In that case the dendrites that do not receive the pulse input merely serve to modify the total somatic impedance. Application of rule I of \citep{Koch1985} allows us to represent these dendrites (indexed by $d'$) as impedances:
\begin{equation}
Z_{d'}(\omega) = \frac{z_{cd'}(\omega)}{\text{tanh}(\gamma_{d'} (\omega) L_{d'})}
\end{equation}
and rule II allows us to modify the somatic impedance as
\begin{equation}
Z_{\text{som}}'(\omega) = \left( \frac{1}{Z_{\text{som}}(\omega)} + \sum_{d'} \frac{1}{Z_{d'}(\omega)}\right)^{-1}
\end{equation}
Thus the entire effect of the rest of the morphology is summarized in the modified lumped-soma boundary condition
\begin{equation}\label{eq:lsmb}
I_{ld}(0,\omega) = \frac{1}{Z_{\text{som}}'(\omega)}V_{\text{som}}(\omega).
\end{equation}
The Green's function in the frequency domain at location $x$ then follows from solving equation \eqref{eq:freqcable} for boundary conditions \eqref{eq:lsmb} and \eqref{eq:bcfreq}, with $I_d(x,t) = \delta(x-x_i)\delta(t)$, and thus $I_d(x,\omega) = \delta(x-x_i)$. From \citep{Butz1974} it follows that
\begin{equation}\label{eq:gf}
\begin{aligned}
 & G_{dd}(x,x_i,\omega) = z_{cd}(\omega)^2 \text{cosh}(\gamma_d(\omega) (L_d-x_i)) \cdot \\ 
 & \hspace{4mm} \frac{\left( \text{sinh}(\gamma_d(\omega) x) + \frac{Z_{\text{som}}'(\omega)}{z_{cd}(\omega)} \text{cosh}(\gamma_d(\omega) x) \right) }{z_c(\omega) \text{sinh}(\gamma_d(\omega) L_d) + Z_{\text{som}}'(\omega) \text{sinh}(\gamma_d(\omega) L_d)},
\end{aligned}
\end{equation}
for $x \leqslant x_i$, where $z_{cd}(\omega) = \frac{z_{ad}}{\gamma_d(\omega)}$ is the characteristic impedance of dendrite $d$. Evaluating this function at $x=x_i$ yields the transfer function of synapse $i$, putting $x=x_j \ (x_j < x_i)$ gives the transfer function between synapse $i$ and synapse $j$ and $x=0$ results in the transfer function between synapse and soma. The Green's function for $x>x_i$ follows from interchanging $x$ and $x_i$ in \eqref{eq:gf}.
To compute the effect of a synaptic input on the driving force in other branches (denoted by $d'$), we first use equation \eqref{eq:gf} (corresponding to rule III of \citep{Koch1985}) to obtain the pulse-voltage response in the frequency domain at the soma. Then, to compute the pulse voltage response in the branch where the driving force needs to be known, we use the following identity:
\begin{equation}
G_{dd'}(x,x',\omega) = \frac{G_{d'd'}(x,0,\omega) G_{dd}(0,x',\omega)}{G(\text{soma}, \text{soma}, \omega)},
\end{equation}
corresponding to rule IV of \citep{Koch1985}.

\subsubsection{Transforming the Green's function to the time domain}

Given the conventions we assumed when transforming the original equation, the inverse Fourier transform has following form:
\begin{equation}\label{eq:transint}
G(x,x_i,t) = \frac{1}{2\pi} \int_{-\infty}^{\infty}\mathrm{d}\omega \ G(x,x_i,\omega) \ e^{i \omega t}.
\end{equation}
If the Green's function in the time-domain rises continuously from zero, which is generally the case if $x \neq x_i$, it can be approximated with negligible error by the standard technique for evaluating Fourier integrals with the fast-Fourier transform (FFT) algorithm \citep{Press2007Numerical}: we choose a sufficiently large interval $[-\omega_m,\omega_m]$ (where $G(x,x_i,\pm \omega_m)$ is practically 0), divide it in $M=2^n$ pieces of with $\Delta \omega = \frac{2\omega_m}{M}$ and approximate the integral by a discrete sum:
\begin{equation}\label{eq:tranform}
G(x,x_i,t) = \frac{1}{2\pi} \sum_{j=0}^{M-1}G(x,x_i,\omega_j)e^{i\omega_j t},
\end{equation}
where $\omega_j = -\omega_m + j \Delta \omega$. The choice of discretization step then fixes the timestep $\Delta t = \frac{2\pi}{M \Delta \omega}$. Upon evaluating the Green's function in the time-domain at $t_l = l \Delta t, \ l=0,\hdots,\frac{M}{2}-1$, expression \eqref{eq:tranform} can be written in a form that is suitable for the fast Fourier transform algorithm:
\begin{equation}
G(x,x_i,t_l) = \frac{\Delta \omega}{2\pi} e^{-i \omega_m t_l} \sum_{j=0}^{M-1}G(x,x_i,\omega_j)e^{i\frac{2\pi}{M}jl},
\end{equation}
and hence:
\begin{equation}
G(x,x_i,t_l) = \frac{M \Delta \omega}{2\pi} e^{-i \omega_m t_l} \text{FFT}(G(x,x_i,\omega_j))_l
\end{equation}
The situation is different if we consider the Green's function at the input location ($x = x_i$). There, the function rises discontinuously from zero at $t=0$, which causes the spectrum in the frequency-domain to have non-vanishing values at arbitrary high frequencies. Hence, the effect of integrating over a finite interval $[-\omega_m,\omega_m]$ will be non-negligible. Formally, this truncation can be interpreted as multiplying the original function with a window function $H(\omega)$ that is 1 in the interval $[-\omega_m,\omega_m]$ and 0 elsewhere, resulting in a time-domain function that is a convolution of the real function and the transform of the window:
\begin{equation}
\begin{aligned}
& \tilde{G}(\omega) = G(x_i,x_i,\omega)H(\omega) \hspace{4mm} \\
& \hspace{4mm} \Longrightarrow \hspace{4mm} \tilde{G}(t) = \int_{-\infty}^{\infty} G(x_i,x_i,\tau)H(t-\tau).
\end{aligned}
\end{equation}
For the rectangular window, the transform $H(t)$ has significant amplitude components for $t\neq 0$, an unwanted property that will cause the Green's function to have spurious oscillations, a phenomenon that is known as spectral leakage \citep{Blackman1958}. This problem can be solved by chosing a different window function, which is 1 at the center of the spectrum and drops continuously to zero at $-\omega_m$ and $\omega_m$. For this work we found that the Hanning window,
\begin{equation}
H(\omega) = \frac{1}{2}\left(1+\cos \left( \frac{\pi \omega}{\omega_m} \right) \right),
\end{equation} 
gave accurate results for $t\neq 0$. For $t=0$, the amplitude is slightly underestimated as a consequence of the truncation of the spectrum, whereas for $t$ very close to, but larger than $0$, the amplitude is slightly overestimated. However, these errors only cause discrepancy in a very small window ($<\unit[0.1]{ms}$) and thus have negligible effect on the neural dynamics.

\section{Model implementation \& Validation}

\subsection{Synapse model implementation}

We implemented a prototype of the synapse model discussed above in two stages. First, after specifying the morphology and the synapse locations, the Green's Function is evaluated at the locations that are needed to solve the system, thus yielding a set of pulse response kernels. As modern high-level languages can handle vectorization very efficiently, these functions can be evaluated for a large set of frequencies $\omega$ quickly, thus allowing for great accuracy. Second, we implemented a model neuron that uses these Green's functions, sampled at the desired temporal accuracy. Then, given a set of synaptic parameters, the somatic membrane potential is computed by integrating the Volterra-equations \eqref{eq:greenssynapse} and \eqref{eq:greenssoma} \citep{Press2007Numerical}. 

\begin{table}[htb]
\centering
\begin{tabular}{|l|l|l|l|l|}
\hline 
\multicolumn{5}{|c|}{Physiology} \\
\hline 
$C_m$ & \multicolumn{4}{c|}{\unit[1]{$\mu F/cm^2$}} \\
$g_m$ & \multicolumn{4}{c|}{\unit[0.02]{$mS/cm^2$}} \\
$r_a$ & \multicolumn{4}{c|}{\unit[100]{$\Omega cm$}} \\
$E_l$ & \multicolumn{4}{c|}{\unit[-65]{mV}} \\
\hline
\multicolumn{5}{|c|}{Morphology} \\
\hline
Soma length & \multicolumn{4}{c|}{\unit[25]{$\mu m$}} \\
Soma diam   & \multicolumn{4}{c|}{\unit[25]{$\mu m$}} \\
\hline
& \multicolumn{2}{|c|}{Fig~\ref{fig:input_order}B} & \multicolumn{2}{c|}{Fig~\ref{fig:input_order}C} \\
\hline
& dend 1 & dend 2 &  dend 1 & dend 2 \\ \cline{2-5}
$L_d$ & \unit[950]{$\mu$m} & \unit[450]{$\mu$m} & \unit[900]{$\mu$m} & \unit[500]{$\mu$m} \\
$a_d$ & \unit[0.25]{$\mu$m} & \unit[0.5]{$\mu$m} & \unit[0.5]{$\mu$m} & \unit[1]{$\mu$m}\\
\hline
\multicolumn{5}{|c|}{Synapses} \\
\hline
& syn 1 & syn 2 & syn 1 & syn 2 \\ \cline{2-5}
$E_r$ & \unit[0]{mV} & \unit[0]{mV} & \unit[0]{mV} & \unit[0]{mV} \\
$\tau$ & \unit[1.5]{ms} & \unit[1.5]{ms} & \unit[1.5]{ms} & \unit[1.5]{ms} \\
$\overline{g}$ & \unit[5]{nS} & \unit[2]{nS} & \unit[20]{nS} & \unit[9]{nS} \\
\hline
\end{tabular}
\caption{Model neuron parameters. The multi-compartmental model explicitly simulates the dendritic structure, while the point-neuron is equipped with our model synapse based on Green's functions and implicitly simulates the dendritic structure.}
\label{table:parameters}
\end{table}

\subsection{Multi-compartmental and point-neuron model}

To compare the performance between a multi-compartmental model and a point-neuron model using the proposed synapse model, we created two comparable neuron models. In the multi-compartmental model, the dendrites are modeled explicitly using \textsc{neuron} \citep{Carnevale2006}, while in the point-neuron model the dendrites are omitted and dendritic processing is carried out implicitly by the new synapse model. The properties of both model neurons are listed in Table~\ref{table:parameters}. Evidently, the implicit model has no real morphology and the parameters related to the geometry are used to instantiate the synapse model. 

\subsection{Input-order detection with differential dendritic filtering}\label{sec:iodetect}

\begin{figure*}[htb!]
   \centering
   \includegraphics[width=0.9\textwidth]{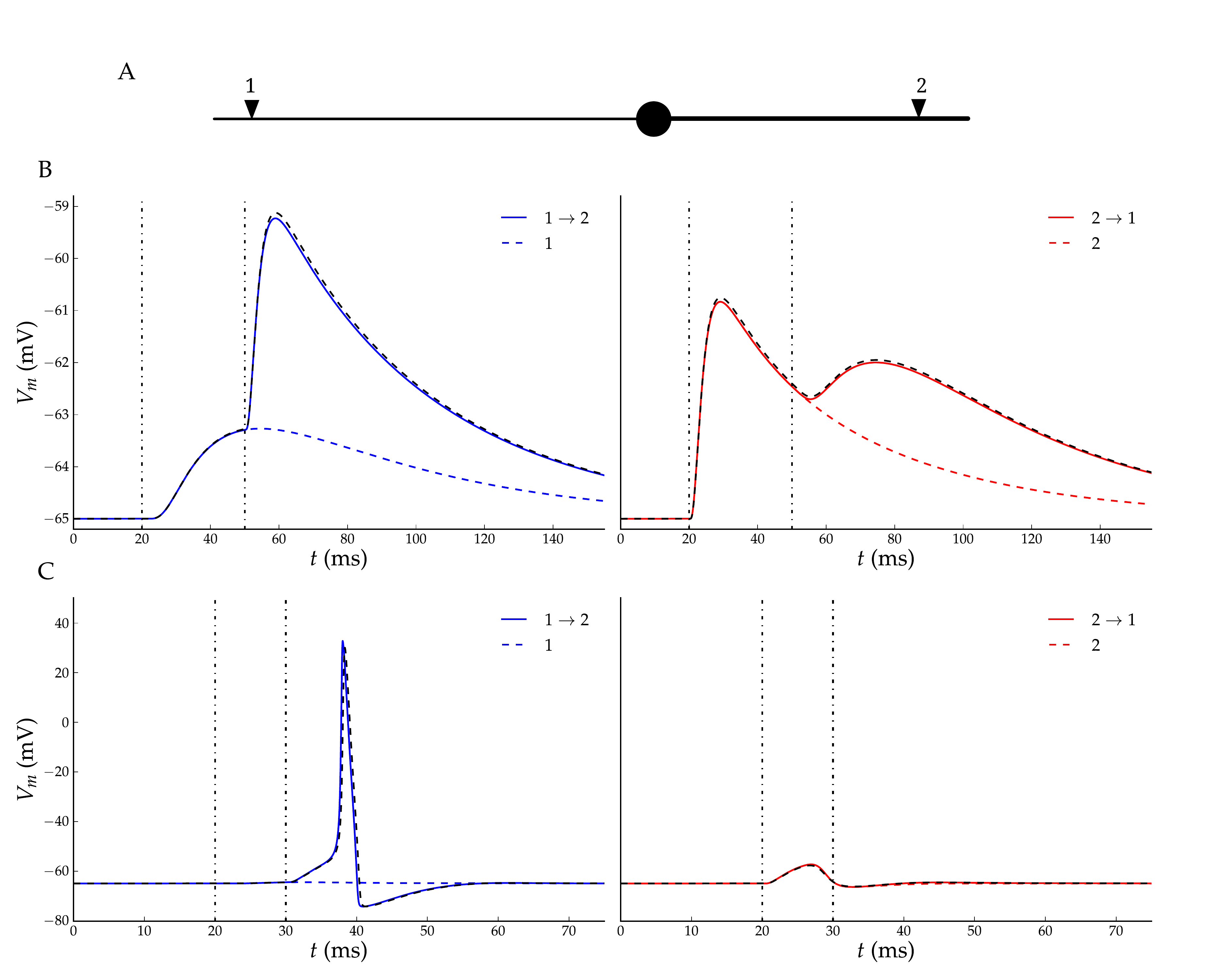} 
   \caption{Comparison  between a reference multi-compartmental model and a point-neuron model equipped with the new synapse model implicetely simulating dendritic processing. A: Both model neurons performed the input-order detection task: The neuron has to respond as strong as possible to the temporal activation 1 $\rightarrow$ 2 and as weak as possible to the reverse temporal order. B: The input-order dectection task for a completely passive neuron. Left and right panels contain the somatic membrane potential when the synapses were activated in the preferred ($1 \rightarrow 2$) and null ($2 \rightarrow 1$) temporal order respectively. Colored lines represent the voltage in the point-neuron model and the black dashed line depicts the \textsc{neuron} trace for comparison. As a reference the waveform when only the first synapse is activated is also shown (left: 1 and right: 2). Vertical dashed-dotted lines denote the spikes arriving at synapse 1 and 2 (left) or 2 and 1 (right). (C) Same as (B), but now the soma contained active HH-currents.  
   }
   \label{fig:input_order}
\end{figure*}

To show the applicability of the new type of model synapse, we use it to perform input-order detection: Suppose a neuron with two dendrites and one synapse (or one group of synapses) on either dendrite (shown in figure~\ref{fig:input_order}A). In the input-order task, the neuron has to generate a strong response to the temporal activation of the synapses $1 \rightarrow 2$, while generating a weak response to the reversed temporal activation $2 \rightarrow 1$. This behavior is achieved by differential dendritic filtering and can thus not be achieved in a straight-forward way by a single-compartmental model. 

We compared the implicit point-neuron model equipped with the new synapse model to the explicit multi-compartmental model in the input-order detection task. The results are illustrated in figure~\ref{fig:input_order}B. Somatic membrane voltages are shown for the point-neuron model and the multi-compartmental model, after synapse activation in the preferred (left) and null temporal order (right). Because the traces are nearly identical, this result validates our approach and the implementation of the synapse model based on the Green's function solution to the cable theory. 

\subsection{Voltage-gated active currents}

The most prominent non-linear neuronal response is the action potential. Since it is possible in our synapse model to include any non-linear conductance mechanism, as long as it is spatially restricted to a point-like location, we built a prototype containing the $\text{Na}^+$ and $\text{K}^+$ conductances required to generated action potentials. By computing the kernels needed to run the upgraded point-neuron model in the input-order detection task and by adjusting the synaptic weights, we yielded a point-neuron model able to generate a spike in response to the preferred activation pattern, while remaining silent in response to the reversed temporal activation. Note that the active somatic currents shorten the timescale of the neuron's response compared to the passive model. The timscale of the t-axis was scaled accordingly. In order to validate these outcomes, we again built an equivalent multi-compartmental model in \textsc{neuron} in which we inserted the same $\text{Na}^+$ and $\text{K}^+$ conductances into the soma. The multi-compartmental model generated identical results, as shown in Figure~\ref{fig:input_order}C. Thus, in principle we can include conductance descriptions to obtain hallmark neuronal non-linearities.

\subsection{Multiple synapse interactions}

\begin{figure*}[htb!]
   \centering
    \includegraphics[width=0.9\textwidth]{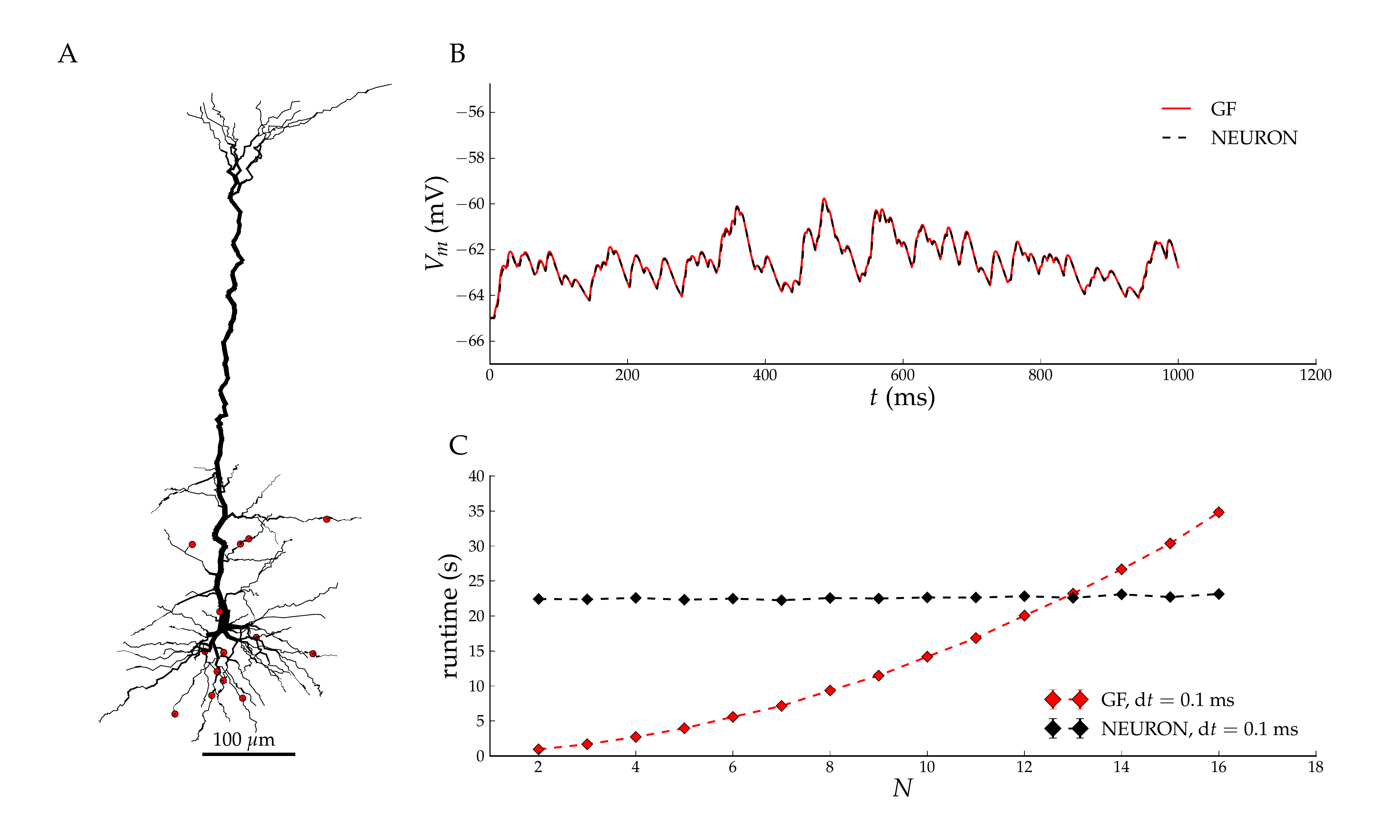}
   \caption{Comparison between the ``implicit'' (red lines) and ``explicit'' (black lines) model neurons of a pyramidal cell stimulated by Poisson spiketrains. A: The neuron morphology together with the synapse locations. B: The membrane potential traces at the soma, for the input locations shown in panel A (red dots). C: Comparison of the runtime versus the number of input locations. For few input locations, our prototype python code outperforms the \textsc{neuron} code.}
   \label{fig:spiketrain}
\end{figure*}

We then checked the correctness of the integrative properties of our implicit point-neuron model by stimulating it with realistic spiketrains at multiple synapses. To that end we added five synapses to a model of a Layer 5 pyramidal neuron equipped with a experimentally reconstructed morphology. The morphology wad retrieved from the NeuroMorpho.org repository \citep{Ascoli2007} and originally published in \citep{Wang2002}. We stimulated each synapse with Poisson spike trains of rate \unit[10]{Hz}. The result is shown in Figure~\ref{fig:spiketrain}. Again, we compared the implicit model's membrane potential traces to the traces obtained from a multi-compartmental model. The agreement is excellent, as can be seen in Figure~\ref{fig:spiketrain}B, which also validates our approach when processing inputs from multiple, interacting synapses. 

\subsection{Runtime}
We established that the ``implicit'' model neuron equipped with our new synapse model generated near-identical voltage traces as a reference multi-compartmental model. Next we compared the run-time of our implementation to the gold standard in multi-compartmental modeling, the \textsc{neuron} software \citep{Carnevale2006}. To this end we simulated a detailed multi-compartmental model (Figure~\ref{fig:spiketrain}) in \textsc{neuron} as well as with our approach, for increasing numbers of input locations. For each of those numbers we ran three simulations of 1 second of simulated time at an integration step of 0.1 ms (10 kHz). Because in our approach the execution time is independent of the morphological complexity but rather scales with the number of input locations, it is expected that for a low number of input locations, applying our model will be much faster. As shown in Figure~\ref{fig:spiketrain} C, for two input locations, our approach runs 20 times faster than \textsc{neuron}, while at 13 input locations the execution time is equal. Keeping in mind \textit{i}) that our implementation is done in Python, and \textit{ii}) that often synapses can be grouped together \citep{Pissadaki2010} we consider this a good outcome.

\section{Discussion}
\label{sec:discussion}

We presented a bridge between single-compartment and multi-compartmental neuron models by creating a synapse model that analytically computes the dendritic processing between the synaptic input locations and the soma. We then demonstrated that point-neuron models equipped with this new synapse model could flawlessly perform the input-order detection computation; a neuronal computation exploiting differential dendritic processing \citep{Agmon-Snir1998}. Thus, the new synapse model can be used to introduce computations to point-neurons that previously only belonged to the realm of multi-compartmental neuron models, with a computational cost that does not depend on the morphological complexity. 

Then the question arises when it would be advisable to use our synapse model over the standard tools. Although a quantitative comparison should be treated with care due to the different implementation languages, we still found that our Python-prototype was much faster than the optimized, C++-based \textsc{neuron}-simulation when the number of input locations was low. This, together with the fact that the computational cost of our model does not depend on morphological complexity, then defines the use case for our model. In scenarios where the number of input locations is low, as is the case in some (invertebrate) cells \citep{Bullock1965} and as in many \emph{in-silico} scenarios, only few Volterra equations have to be integrated. There our model represents considerable computational advantage. This arguments also holds when more complex neuron types are considered: while cortical neurons receive often as many as 10000 synapses, many of those can be grouped together. To a good approximation, small dendritic branches act as single units, both in terms of short-term input integration \citep{Poirazi2003, London2005} as in terms of long-term plasticity related processes \citep{Govindarajan2011}. Thus, one could group all synapses in a small branch together and then compute the Green's function for that group of synapses as a hole. Such a grouping would drastically reduce the number of Volterra equations to be integrated and hence enhance performance accordingly. 


We assumed that the PSP waveform is transformed only in a passive manner on its way to the soma. In reality, this might sound like a drastic simplification as non-linearity is often cited as a hallmark of neuronal computation, not in the least to generate output spikes. How can we evaluate our synapse model in the light of non-linear computations?

Non-linearities in neural response can occur in two ways. First, at the synapse level a non-linear response can be generated principally through the recruitment of NMDA receptors during repetitive synaptic activation \citep{Branco2010}. As we assume the evolution in time of the synaptic conductance to be of a known shape, we could -in principle- also mimic a non-linear synaptic conductance by using a more specific description of the synaptic conductance evolution.

Second, non-linearities can arise from voltage-gated conductances in neuronal membranes, that are often distributed non-uniformly along the dendrite \citep{Larkum1999, Angelo2007, Mathews2010}. The distributed nature of voltage-gated conductances leads to the view that dendritic processing is non-linear, and shaped by these conductances and their spatial distributions. Recent work actually challenges this view as it is known that in some behavioral regimes, dendrites act linearly \citep{Ulrich2002, Schoen2012}. Since our Green's function approach relies only on the assumption of linearity, it is not intrinsically restricted to passive dendrites. Ion channels distributed along a dendrite can be linearized \citep{Mauro1970}, and thus yield a quasi-active cable \citep{Koch1998}. We anticipate that such a linearization procedure can be plugged into our synapse model, so that the linear (but active) properties of the membrane are captured in the Green's function, yielding accurate and efficient simulations of dendrites that reside in their linear regime. Also, in some cases the actual distribution of voltage-gated conductances along the dendrite does not seem to have any effect as long as the time constant for activation is slower than the spread of voltage itself, which makes the actual location of the voltage-gated conductance irrelevant \citep{Angelo2007}. Thus, in those cases were the spread of voltage is faster than the activation of the conductance, dendrites can act in a passive way, as long as the appropriate non-linearity is introduced at one or a few point-like locations. This can be introduced easily in our synapse model (see Figure~\ref{fig:input_order}C, with the soma as point-like location with active currents).

While dealing with neuronal non-linearities the focus is often on supra-linear responses to inputs, despite the fact that sub-linear responses are also intrinsically non-linear. Moreover, recently it has been shown both in theory and experiment that sub-linear response are used by neurons \citep{Vervaeke2012,Abrahamsson2012}. Even in passive dendrites, sub-linear responses can be generated when the dendrite locally saturates: due to high input resistance the local voltage response to an input can reach the reversal potential of the membrane. At that moment the driving force disappears and a sub-linear response is generated to inputs. This sort of sub-linear response can be generated in conductance-based models with realistic morphologies. Because we implicitly model dendritic morphology, our synapse model is capable of generating these sub-linear responses.

In conclusion, we presented a new synapse model that computes the PSP waveforms as if they were subject to dendritic processing without the need to explicitly simulate the dendrites themselves. With this synapse model comes the ability to simulate dendritic processing at a low computational complexity, that allows it's incorporation in large scale models of neural networks. We thus made a first step to bridge single and multi-compartmental modeling.

\subsubsection*{Acknowledgements}
We thank Marc-Oliver Gewaltig for comments on the manuscript and Moritz Deger for helpful discussion. This work was supported by the BrainScaleS EU FET-proactive FP7 grant.


\begin{thebibliography}{}

\bibitem[Abrahamsson et~al., 2012]{Abrahamsson2012}
Abrahamsson, T., Cathala, L., Matsui, K., Shigemoto, R., and Digregorio, D.~a.
  (2012).
\newblock {Thin dendrites of cerebellar interneurons confer sublinear synaptic
  integration and a gradient of short-term plasticity.}
\newblock {\em Neuron}, 73(6):1159--72.

\bibitem[Agmon-Snir et~al., 1998]{Agmon-Snir1998}
Agmon-Snir, H., Carr, C.~E., and Rinzel, J. (1998).
\newblock {The role of dendrites in auditory coincidence detection}.
\newblock {\em Nature}, 393(May).

\bibitem[Angelo et~al., 2007]{Angelo2007}
Angelo, K., London, M., Christensen, S.~R., and H\"{a}usser, M. (2007).
\newblock {Local and global effects of I(h) distribution in dendrites of
  mammalian neurons.}
\newblock {\em The Journal of neuroscience : the official journal of the
  Society for Neuroscience}, 27(32):8643--53.

\bibitem[Ascoli et~al., 2007]{Ascoli2007}
Ascoli, G.~A., Donohue, D.~E., and Halavi, M. (2007).
\newblock {NeuroMorpho.Org: A Central Resource for Neuronal Morphologies}.
\newblock {\em The Journal of Neuroscience}, 27(35):9247--9251.

\bibitem[Blackman and Tukey, 1958]{Blackman1958}
Blackman, R. and Tukey, J. (1958).
\newblock {\em {The measurement of power spectra}}.
\newblock Dover publications.

\bibitem[Branco et~al., 2010]{Branco2010}
Branco, T., Clark, B.~A., and H\"{a}usser, M. (2010).
\newblock {Dendritic discrimination of temporal input sequences in cortical
  neurons.}
\newblock {\em Science (New York, N.Y.)}, 329(5999):1671--5.

\bibitem[Brette et~al., 2007]{Brette2007}
Brette, R., Rudolph, M., Carnevale, T., Hines, M., Beeman, D., Bower, J.~M.,
  Diesmann, M., Morrison, A., Goodman, P.~H., Harris, F.~C., Zirpe, M.,
  Natschl\"{a}ger, T., Pecevski, D., Ermentrout, B., Djurfeldt, M., Lansner,
  A., Rochel, O., Vieville, T., Muller, E., Davison, A.~P., {El Boustani}, S.,
  and Destexhe, A. (2007).
\newblock {Simulation of networks of spiking neurons: a review of tools and
  strategies.}
\newblock {\em Journal of computational neuroscience}, 23(3):349--98.

\bibitem[Bullock and Horridge, 1965]{Bullock1965}
Bullock, T.~H. and Horridge, G.~A. (1965).
\newblock {\em Structure and function in the nervous systems of invertebrates /
  [by] Theodore Holmes Bullock and G. Adrian Horridge. With chapters by Howard
  A. Bern, Irvine R. Hagadorn [and] J. E. Smith}.
\newblock W. H. Freeman San Francisco.

\bibitem[Butz and Cowan, 1974]{Butz1974}
Butz, E.~G. and Cowan, J.~D. (1974).
\newblock {Transient potentials in dendritic systems of arbitrary geometry}.
\newblock {\em Biophysical journal}, 14:661--689.

\bibitem[Carnevale and Hines, 2006]{Carnevale2006}
Carnevale, N.~T. and Hines, M.~L. (2006).
\newblock {\em {The NEURON Book}}.
\newblock Cambridge University Press, New York, NY, USA.

\bibitem[Gewaltig and Diesmann, 2007]{Gewaltig2007}
Gewaltig, M.-O. and Diesmann, M. (2007).
\newblock {NEST (NEural Simulation Tool)}.
\newblock {\em Scholarpedia}, 2(4):1430.

\bibitem[Gidon and Segev, 2012]{Gidon2012}
Gidon, A. and Segev, I. (2012).
\newblock {Principles governing the operation of synaptic inhibition in
  dendrites.}
\newblock {\em Neuron}, 75(2):330--41.

\bibitem[Giugliano, 2000]{Giugliano2000}
Giugliano, M. (2000).
\newblock {Synthesis of generalized algorithms for the fast computation of
  synaptic conductances with markov kinetic models in large network
  simulations}.
\newblock {\em Neural Computation}, 931:903--931.

\bibitem[Govindarajan et~al., 2011]{Govindarajan2011}
Govindarajan, A., Israely, I., Huang, S.-Y., and Tonegawa, S. (2011).
\newblock {The dendritic branch is the preferred integrative unit for protein
  synthesis-dependent LTP.}
\newblock {\em Neuron}, 69(1):132--46.

\bibitem[G\"{u}tig and Sompolinsky, 2006]{Gutig2006}
G\"{u}tig, R. and Sompolinsky, H. (2006).
\newblock {The tempotron: a neuron that learns spike timing-based decisions.}
\newblock {\em Nature neuroscience}, 9(3):420--8.

\bibitem[Hay et~al., 2013]{Hay2013}
Hay, E., Sch\"{u}rmann, F., Markram, H., and Segev, I. (2013).
\newblock {Preserving Axo-somatic Spiking Features Despite Diverse Dendritic
  Morphology.}
\newblock {\em Journal of neurophysiology}.

\bibitem[Jolivet et~al., 2004]{Jolivet2004}
Jolivet, R., Lewis, T.~J., and Gerstner, W. (2004).
\newblock {Generalized integrate-and-fire models of neuronal activity
  approximate spike trains of a detailed model to a high degree of accuracy.}
\newblock {\em Journal of neurophysiology}, 92(2):959--76.

\bibitem[Kellems et~al., 2010]{Kellems2010}
Kellems, A.~R., Chaturantabut, S., Sorensen, D.~C., and Cox, S.~J. (2010).
\newblock {Morphologically accurate reduced order modeling of spiking neurons.}
\newblock {\em Journal of computational neuroscience}, 28(3):477--94.

\bibitem[Koch, 1998]{Koch1998}
Koch, C. (1998).
\newblock {\em {Biophysics of Computation: Information Processing in Single
  Neurons (Computational Neuroscience)}}.
\newblock Oxford University Press, 1 edition.

\bibitem[Koch and Poggio, 1985]{Koch1985}
Koch, C. and Poggio, T. (1985).
\newblock {A simple algorithm for solving the cable equation in dendritic trees
  of arbitrary geometry}.
\newblock {\em Journal of neuroscience methods}.

\bibitem[Larkum et~al., 1999]{Larkum1999}
Larkum, M.~E., Zhu, J.~J., and Sakmann, B. (1999).
\newblock {A new cellular mechanism for coupling inputs arriving at different
  cortical layers.}
\newblock {\em Nature}, 398(6725):338--41.

\bibitem[London and H\"{a}usser, 2005]{London2005}
London, M. and H\"{a}usser, M. (2005).
\newblock {Dendritic computation.}
\newblock {\em Annual review of neuroscience}, 28:503--32.

\bibitem[Magee, 1999]{Magee1999}
Magee, J.~C. (1999).
\newblock {Dendritic Ih normalizes temporal summation in hippocampal CA1
  neurons}.
\newblock {\em Nature neuroscience}, 2(9):848.

\bibitem[Markram, 2006]{Markram2006}
Markram, H. (2006).
\newblock {The blue brain project.}
\newblock {\em Nature reviews. Neuroscience}, 7(2):153--60.

\bibitem[Mathews et~al., 2010]{Mathews2010}
Mathews, P.~J., Jercog, P.~E., Rinzel, J., Scott, L.~L., and Golding, N.~L.
  (2010).
\newblock {Control of submillisecond synaptic timing in binaural coincidence
  detectors by K(v)1 channels.}
\newblock {\em Nature neuroscience}, 13(5):601--9.

\bibitem[Mauro et~al., 1970]{Mauro1970}
Mauro, A., Conti, F., Dodge, F., and Schor, R. (1970).
\newblock {Subthreshold behavior and phenomenological impedance of the squid
  giant axon.}
\newblock {\em The Journal of general physiology}, 55(4):497--523.

\bibitem[Migliore and Shepherd, 2002]{Migliore2002}
Migliore, M. and Shepherd, G.~M. (2002).
\newblock {Emerging rules for the distributions of active dendritic
  conductances.}
\newblock {\em Nature reviews. Neuroscience}, 3(5):362--70.

\bibitem[Norman, 1972]{Norman1972}
Norman, R.~S. (1972).
\newblock {Cable theory for finite length dendritic cylinders with initial and
  boundary conditions}.
\newblock {\em Biophysical journal}, 12(1):25--45.

\bibitem[Ohme and Schierwagen, 1998]{Ohme1998}
Ohme, M. and Schierwagen, A. (1998).
\newblock {An equivalent cable model for neuronal trees with active membrane.}
\newblock {\em Biological cybernetics}, 78(3):227--43.

\bibitem[Pissadaki et~al., 2010]{Pissadaki2010}
Pissadaki, E.~K., Sidiropoulou, K., Reczko, M., and Poirazi, P. (2010).
\newblock {Encoding of spatio-temporal input characteristics by a CA1 pyramidal
  neuron model.}
\newblock {\em PLoS computational biology}, 6(12):e1001038.

\bibitem[Poirazi et~al., 2003]{Poirazi2003}
Poirazi, P., Brannon, T., and Mel, B.~W. (2003).
\newblock {Pyramidal neuron as two-layer neural network}.
\newblock {\em Neuron}, 37:989--999.

\bibitem[Press et~al., 2007]{Press2007Numerical}
Press, W.~H., Teukolsky, S.~A., Vetterling, W.~T., and Flannery, B.~P. (2007).
\newblock {\em {Numerical Recipes 3rd Edition: The Art of Scientific
  Computing}}.
\newblock Cambridge University Press, New York, NY, USA, 3 edition.

\bibitem[Richert et~al., 2011]{Richert2011}
Richert, M., Nageswaran, J.~M., Dutt, N., and Krichmar, J.~L. (2011).
\newblock {An efficient simulation environment for modeling large-scale
  cortical processing.}
\newblock {\em Frontiers in neuroinformatics}, 5(September):19.

\bibitem[Rotter and Diesmann, 1999]{Rotter1999}
Rotter, S. and Diesmann, M. (1999).
\newblock {Exact digital simulation of time-invariant linear systems with
  applications to neuronal modeling.}
\newblock {\em Biological cybernetics}, 81(5-6):381--402.

\bibitem[Schoen et~al., 2012]{Schoen2012}
Schoen, A., Salehiomran, A., Larkum, M.~E., and Cook, E.~P. (2012).
\newblock {A compartmental model of linear resonance and signal transfer in
  dendrites.}
\newblock {\em Neural computation}, 24(12):3126--44.

\bibitem[Spruston, 2008]{Spruston2008}
Spruston, N. (2008).
\newblock {Pyramidal neurons: dendritic structure and synaptic integration}.
\newblock {\em Nature Reviews Neuroscience}, 9(3):206--221.

\bibitem[Torben-Nielsen and Stiefel, 2010]{Torben-Nielsen2010}
Torben-Nielsen, B. and Stiefel, K.~M. (2010).
\newblock {An inverse approach for elucidating dendritic function.}
\newblock {\em Frontiers in computational neuroscience}, 4(September):128.

\bibitem[Traub et~al., 2005]{Traub2005}
Traub, R.~D., Contreras, D., Cunningham, M.~O., Murray, H., LeBeau, F. E.~N.,
  Roopun, A., Bibbig, A., Wilent, W.~B., Higley, M.~J., and Whittington, M.~a.
  (2005).
\newblock {Single-column thalamocortical network model exhibiting gamma
  oscillations, sleep spindles, and epileptogenic bursts.}
\newblock {\em Journal of neurophysiology}, 93(4):2194--232.

\bibitem[Tuckwell, 1988]{Tuckwell1988Introduction}
Tuckwell, H.~C. (1988).
\newblock {\em {Introduction to theoretical neurobiology}}.
\newblock Cambridge studies in mathematical biology, 8. Cambridge University
  Press.

\bibitem[Ulrich, 2002]{Ulrich2002}
Ulrich, D. (2002).
\newblock {Dendritic Resonance in Rat Neocortical Pyramidal Cells}.
\newblock {\em Journal of Neurophysiology}, 87(6):2753--2759.

\bibitem[{Van Pelt}, 1992]{VanPelt1992}
{Van Pelt}, J. (1992).
\newblock {A simple vector implementation of the Laplace-transformed cable
  equations in passive dendritic trees}.
\newblock {\em Biological cybernetics}, 21:15--21.

\bibitem[Vervaeke et~al., 2012]{Vervaeke2012}
Vervaeke, K., Lorincz, A., Nusser, Z., and Silver, R.~A. (2012).
\newblock {Gap Junctions Compensate for Sublinear Dendritic Integration in an
  Inhibitory Network.}
\newblock {\em Science (New York, N.Y.)}, 1624.

\bibitem[Wang et~al., 2002]{Wang2002}
Wang, Y., Gupta, A., Toledo-Rodriguez, M., Wu, C.~Z., and Markram, H. (2002).
\newblock {Anatomical, physiological, molecular and circuit properties of nest
  basket cells in the developing somatosensory cortex.}
\newblock {\em Cerebral cortex (New York, N.Y. : 1991)}, 12(4):395--410.

\end{thebibliography}

\end{document}